\documentclass[aps,pra,twocolumn,10pt,bibnotes]{revtex4-1} 
\usepackage{graphicx}
\usepackage{amsmath,ulem}
\usepackage{color}
\usepackage{lineno}
\usepackage{epstopdf}

\def\Rb{$^{87}$Rb }
\def\rtf{r_{\mbox{\tiny\itshape TF}}}

\def\ket#1{\left|#1\right\rangle}

\begin{document}
\title{Spontaneous symmetry breaking in spinor Bose-Einstein condensates}

\author{M.~Scherer$^1$, B.~L\"ucke$^1$, O.~Topic$^1$, G.~Gebreyesus$^2$, F.~Deuretzbacher$^2$, W.~Ertmer$^1$, L.~Santos$^2$,  C.~Klempt$^1$, J.J.~Arlt$^3$}

\affiliation{$^1$ Institut f\"ur Quantenoptik, Leibniz Universit\"at Hannover, Welfengarten~1, D-30167~Hannover, Germany}
\affiliation{$^2$ Institut f\"ur Theoretische Physik, Leibniz Universit\"at Hannover, Appelstra\ss e~2, D-30167~Hannover, Germany}
\affiliation{$^3$ QUANTOP, Danish National Research Foundation Center for Quantum Optics, Institut for Fysik og Astronomi, Aarhus Universitet, Ny Munkegade 120, DK-8000 Aarhus C, Denmark}

\date{\today}

\begin{abstract}
We present an analytical model for the theoretical analysis of spin dynamics and spontaneous symmetry breaking in a spinor Bose-Einstein condensate (BEC). This allows for an excellent intuitive understanding of the processes and provides good quantitative agreement with experimental results in Ref.~\cite{Scherer2010}. It is shown that the dynamics of a spinor BEC initially prepared in an unstable Zeeman state $m_F=0$~($\ket{0}$) can be understood by approximating the effective trapping potential for the state $\ket{\pm 1}$ with a cylindrical box potential. The resonances in the creation efficiency of these atom pairs can be traced back to excitation modes of this confinement. The understanding of these excitation modes allows for a detailed characterization of the symmetry breaking mechanism, showing how a twofold spontaneous breaking of spatial and spin symmetry can occur. In addition a detailed account of the experimental methods for the preparation and analysis of spinor quantum gases is given.
\end{abstract}

\maketitle

\section{Introduction}

Spontaneous symmetry breaking is a fundamental process that plays a key role in many fields of physics~\cite{Huang1987}. In particular, it appears in physical scenarios ranging from cosmology~\cite{Kibble1976} and particle physics~\cite{Turok1990} to liquid crystals~\cite{Chuang1991} and superfluid Helium~\cite{Zurek1985}. In these scenarios, small fluctuations typically break some symmetry of the system, and thus determine its dynamical evolution and final state. In particular, this can lead to final states that do not reflect the underlying symmetry of the dynamics because the symmetric state is unstable.

A number of recent experiments have shown that Bose-Einstein condensates (BEC) can provide unprecedented possibilities to study symmetry breaking processes~\cite{Saito2007a}. In particular, the ability to investigate non-equilibrium dynamics~\cite{Polkovnikov2011} including the formation of topological defects via the Kibble-Zurek mechanism~\cite{Kibble1976,Zurek1985} allows for the detailed analysis of dynamical symmetry breaking. The experimental realization are relatively diverse and include vortex formation~\cite{Weiler2008}, spinor BECs~\cite{Ho1998}, BECs with dipolar interaction~\cite{Vengalattore2008} as well as BECs coupled to an optical cavity \cite{Baumann2011}. Moreover symmetry breaking is crucial in understanding Bose-Einstein condensates and their coherence properties~\cite{Javanainen1996,Cirac1996}.  

In particular spinor Bose-Einstein condensates~\cite{Stamper-Kurn2012}, formed from multiple spin components of a given species, offer fascinating opportunities to analyze spontaneous symmetry breaking. Symmetry breaking was first observed in a spinor BEC quenched from a polar into a ferromagnetic phase~\cite{Sadler2006}. During the subsequent dynamics, ferromagnetic domains and topogical defects were observed in the transverse magnetization, whereas the longitudinal magnetization remained negligible. These experiments provided a major insight in the formation of topological defects, however the effects of the external trapping potential were not investigated. A subsequent experiment investigated the decay of an initial spin texture into a domain structure~\cite{Vengalattore2008}. Later experiments investigated the spontaneous formation of patterns in 1D~\cite{Kronjager2010} and provided a detailed understanding of the spatial and spin symmetry breaking processes~\cite{Scherer2010}.

The theoretical investigation of spinor BEC and of symmetry breaking processes therein is of ongoing interest~\cite{Saito2007a}. Prominent examples include the formation of spin structures breaking the chiral symmetry in spinor condensates with ferromagnetic interactions~\cite{Saito2006}, symmetry breaking in a double-well potential~\cite{Mayteevarunyoo2008} and the relevance of thermal atoms for spontaneous magnetization~\cite{Phuc2011}.

Within our work the particularly interesting case of spinor BECs initially prepared in an unstable $m_F=0$~($\ket{0}$) Zeeman state is investigated. In this case spin changing collisions lead to the creation of correlated atom pairs in $m_F=\pm 1$~($\ket{\pm 1}$) in a process equivalent to parametric down-conversion in nonlinear optics~\cite{Klempt2010}. Resonances in the creation efficiency of  these atom pairs can be traced back to specific excitation modes of the effective confinement~\cite{Klempt2009}. The understanding of these excitation modes allows for a detailed characterization of the symmetry breaking mechanism~\cite{Scherer2010}. It was shown that a twofold spontaneous breaking of spatial and spin symmetry in the amplified $m_F=\pm 1$ clouds can occur.

Here, we present an analytical model for the theoretical analysis of spin dynamics and symmetry breaking in our system. We show that an excellent intuitive understanding and good quantitative agreement with experimental results can be obtained by approximating the effective trapping potential with a cylindrical box. This method is used to provide a detailed analysis of the spin dynamics and symmetry breaking processes in Ref.~\cite{Scherer2010}. In addition a detailed account of the experimental methods and the analysis techniques is provided.

The paper is structured as follows, section~\ref{experiment} describes the production of quantum gases and the experimental techniques used to prepare, investigate and detect spinor BECs. Section~\ref{theory} introduces a theoretical analysis of the system in a simplified box potential. This provides the basis for an understanding of the spontaneous breaking of the spatial and the longitudinal spin symmetry.  Section~\ref{SymmetryBreaking} highlights the experimental results on spontaneous symmetry breaking in view of the occupation of higher spatial modes. 

\section{Experimental measurement of spin dynamics} 
\label{experiment}

The excitation modes within a spinor condensate and the associated rate of spin changing collisions depend strongly on the confining potential and the initial spin population. In the following, the experimental production of spinor gases and their analysis is therefore described in detail.

\subsection{Production of quantum gases}

Initially $5 \times 10^9$~\Rb atoms are loaded from the background vapor into a magneto-optical trap, using light induced atom desorption~\cite{Klempt2006}. After molasses cooling and optical pumping into the low-field seeking state $\ket{F=2,m_F=2}\equiv\ket{2,2}$, the atoms are transferred to a magnetic quadrupole trap and mechanically transported into the experiment chamber formed by a glass cell at ultra-high vacuum. In the next step the atoms are loaded into a harmonic magnetic trap in QUIC configuration~\cite{Esslinger1998} with trap frequencies of $2\pi \times 230$ Hz ($2\pi \times 23$ Hz) in radial (axial) direction, where they are cooled by forced radio-frequency evaporation. This evaporation is stopped shortly before reaching quantum degeneracy and the atoms are transfered~\cite{Klempt2008} into a crossed beam dipole trap (see section~\ref{dipole}) which allows trapping of all Zeeman states. The atoms are further evaporated by lowering the intensities of the two beams until pure Bose-Einstein condensates (BEC) of $5\times 10^4$ atoms in the state $\ket{2,2}$ are reached. To initiate spin dynamics, the atoms are transferred to the state $\ket{2,0}$ (see section~\ref{stateprep}) and the intensities of the dipole trap beams are changed to create the desired trapping potential (see section~\ref{dipole}).

\subsection{Dipole trap}
\label{dipole}

\begin{figure}[ht]
\centering
\includegraphics*[width=0.85\columnwidth]{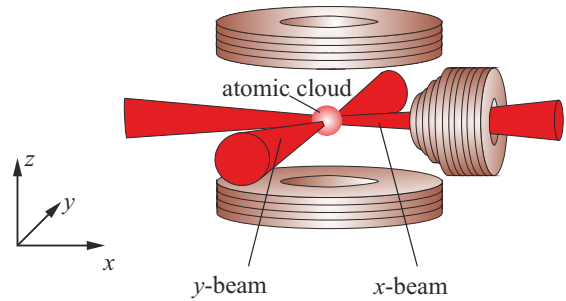}
\caption{Sketch of the crossed-beam dipole trap. The two beams are oriented horizontally and perpendicular to each other. The intersection of the two beams is located at the center of the quadrupole-coils of the QUIC trap~\cite{Esslinger1998}.}
\label{trap}
\end{figure}

The experiments are performed in a red-detuned, crossed-beam dipole trap at a wavelength of $1064$~nm. The linearly polarized light provides an attractive potential, which is independent of the spin state of the atoms. The two beams are aligned perpendicular to each other in the horizontal plane with waists of $54 \mu$m (beam in $x$-direction) and $28\mu$m (beam in $y$-direction) as shown in Fig.~\ref{trap}. 
Depending on the power $P_x$ and $P_y$ in these beams, a variety of trapping configurations can be produced. In all configurations, the lowest trap frequency $\omega_y$ is realized in the $y$-direction, since only the weakly focused beam in $x$-direction significantly contributes to it. For a given power $P_y$, three regimes can be identified depending on the power $P_x$. For low powers $P_x$, a trap with the strongest direction parallel to gravity ($\omega_z >\omega_x $) is realized, while high powers $P_x$ lead to a trap with the highest frequency perpendicular to gravity ($\omega_x >\omega_z $). In the following, these configurations are jointly referred to as elliptical traps. Importantly, there is an intermediate power $P_x$ where a trap with nearly equal trapping frequencies $\omega_z =\omega_x $ can be realized. This is referred to as a cylindrical trap. 

The realization of the trapping configurations outlined above is very sensitive to misalignments and to small astigmatisms of the two beams. In particular these effects can lead to a rotation of the principal axes of the elliptical trap in the $xz$-plane as a function of the relative powers $P_x$ and $P_y$. If these effects are sufficiently large it is indeed not possible to realize a cylindrical trap at all. Nevertheless, this effect can be useful to implement an elliptical trap whose strongest axis has an arbitrary adjustable angle relative to gravity.

It is hence necessary to determine both the trap frequencies and the orientation of the principal axes to adjust the desired trapping potential. To identify these quantities we rely on the center of mass oscillation of the distribution, which is initiated by displacing a BEC in the harmonic potential \cite{Jin1996}. Since this displacement is not necessarily parallel to one of the principal axes, it results in an oscillation along multiple trap axes. After various oscillation times and a fixed free evolution in time-of-flight (TOF) the $x$- and the $z$-position of the clouds are detected by taking absorption images along the $y$-axis. These positions reflect the velocities of the cloud in these directions.

\begin{figure}[ht]
\centering
\includegraphics{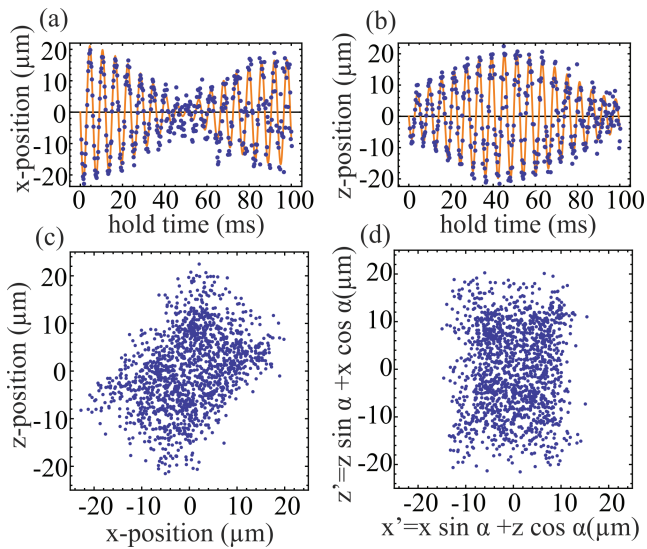}
\caption{Trap frequency measurement in an elliptical trap. (a-b) Positions of the oscillating atomic cloud after release from the dipole-potential. The line is a fit to the data based on two oscillations with mixing angle $\alpha$. (c) Plotting the $x$- and $y$-position against each other results in a tilted Lissajous-like figure. The tilting angle is equivalent to mixing angle obtained from the fit to the data. (d) A principal axes transformation results in a rectangle, which is parallel to the coordinate axes.}
\label{trapfreq1}
\end{figure}

Figure~\ref{trapfreq1}~(a-b) shows the recorded positions. Due to the projection of the trap axes onto the CCD camera axes, a beat signal of two overlapping damped oscillations is observed. If the two positions are plotted against each other as shown in Fig.~\ref{trapfreq1}~(c), this corresponds to a Lissajous figure bounded by a rotated rectangle. To extract the oscillation frequencies and the directions of the principal axes, a superposition of two independent oscillations with mixing angle $\alpha$ is fitted to the data. A principal axes transformation to a frame rotated by $\alpha$ then leads to rectangle parallel to the coordinate axes in the $xz$-plane as shown in Fig.~\ref{trapfreq1}~(d).

\begin{figure}[ht]
\centering
\includegraphics{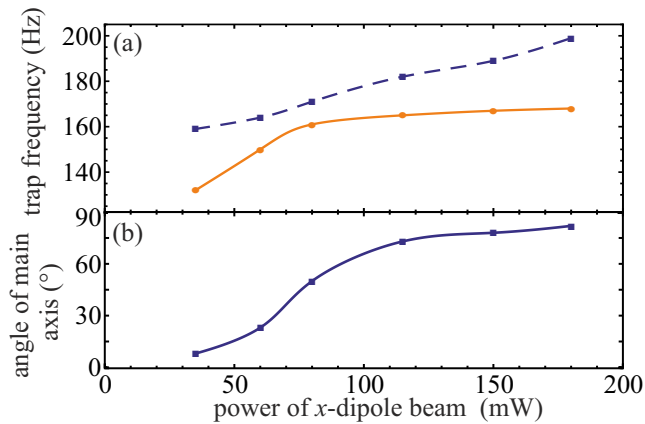}
\caption{Dependence of the trap parameters on the power of the beam in $x$-direction for a trap with a deliberate misalignment of the two beams. $P_x$ is varied from 35~mW to 180~mW while $P_y$ is kept constant at 24~mW. (a) Intermediate (dots, solid line) and strong trap frequency (squares, dashed lane). (b) Angle of the principal axis. All lines are splines to guide the eye.} 
\label{trapfreq2}
\end{figure}

Figure~\ref{trapfreq2} shows the result of such a measurement for a dipole trap with a deliberate misalignment of the two beams. When $P_x$ is increased while $P_y$ is kept fixed, both strong trap frequencies rise and the trap rotates by almost $90\,^{\circ}$. Note however, that the two trap frequencies never become identical and therefore this alignment does not allow for the realization of a cylindrical trap.

The following steps are typically taken to achieve sufficiently good optical alignment to obtain a cylindrical trap. We first measure the aspect ratio and orientation of a BEC after a long TOF and adjust the relative alignment of the beams and the positions of the focusing lenses. Since the expansion of the BEC is closely related to the trap's strength and orientation, these measurements provide a first indication of its geometry. In a second step the trap frequencies are measured as a function of $P_x$ as described above. The cylindrical configuration we typically achieve has trap frequencies of $187$~Hz, $183$~Hz and $67$~Hz and the remaining radial asymmetry is primarily caused by uncompensated astigmatism.

\subsection{State preparation}
\label{stateprep}
	
The preparation of pure spin states is of fundamental importance for two reasons. Firstly, the parametric amplification process investigated here starts in the state $\ket{2,0}$ and is density dependent. Therefore as many atoms as possible have to be transferred from the initial state $\ket{2,2}$ to the state $\ket{2,0}$. Moreover, it is important for the investigation of spontaneous symmetry breaking that the parametric amplification is triggered by quantum fluctuations~\cite{Klempt2010}. Hence the atoms have to be transferred without populating any other spin components, which could act as a spurious seed. 

Starting from the state $\ket{2,2}$, two strategies can be employed for the transfer. One option is a rapid adiabatic passage using radio frequency (RF) radiation. This method has however a number of disadvantages, since it requires relatively high magnetic fields ($\approx 80$~G) to selectively address the individual Zeeman states. 

A more favorable strategy employs two microwave rapid adiabatic passages ($\ket{2,2} \rightarrow \ket{1,1} \rightarrow \ket{2,0}$) to transfer the atoms into the desired state. This approach only requires low magnetic fields ($\approx 6$~G), reducing the required magnetic ramp time. In combination with the absence of hyperfine state changing collisions in the state $\ket{1,1}$ and a fast microwave sweep time (5~ms) it thus allows for a larger initial number of $5\times 10^4$ atoms. Moreover, this method reduces the risk of producing seed atoms in the states $\ket{2,\pm 1}$, since it does not pass any of these during the adiabatic passage. 

In addition to this technique, a strong magnetic field gradient of $58.5$~G/cm is applied for 15~ms after the spin preparation to remove any residual atoms in other states. We have checked the efficiency of this purification method by applying it to BECs of $10^5$~atoms prepared in either one of the states $\ket{2,\pm 1}$. Since no atoms were observed within the detection limit of 500~atoms, a lower limit for the removal efficiency is $99.5~\%$. Hence we estimate that no more than 2.5~atoms remain in the wrong spin state after the preparation sequence.

\subsection{Spin dynamics and detection}
\label{spinexperiment}

To initiate spin dynamics, the following experimental steps are taken. During the purification step the power $P_y$ is ramped to 24~mW whereas $P_x$ is adjusted to a value between 35~mW and 180~mW to realize the desired trap configuration. Subsequently, the direction of the applied homogeneous magnetic field is rotated into the $y$-direction and lowered to a desired value between 0.12~G and 2.5~G in 3~ms. This magnetic field direction is perpendicular to the two strong trap axes. The atoms are then held in the trap for a time of 15~ms to 21~ms to allow for spin changing collisions. 

\begin{figure}[ht]
\centering
\includegraphics{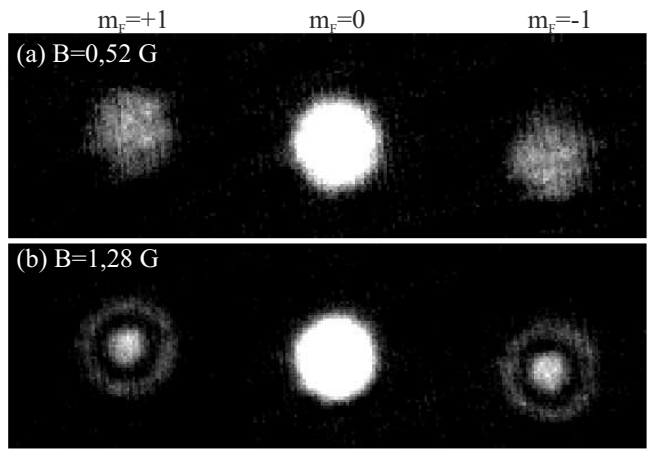}
\caption{Typical absorption images after spin dynamics at two magnetic fields. The spin components were separated by an inhomogeneous magnetic field. The shape of the clouds in the states $\ket{2,\pm 1}$ clearly depends on the applied magnetic field and reflect the spin excitation modes (see section~\ref{spinorcylinder}).}
\label{components}
\end{figure}

Finally, spin dynamics is stopped by switching off the trapping beams. During the following TOF evolution the atomic clouds expand self similarly \cite{Castin1996,Lundh1998,Castin1999,Dalfovo2000,Madison2000}. A strong magnetic field gradient of 37~G/cm is applied in $z$-direction for $3.5$~ms to spatially separate the spin components (Stern-Gerlach technique). After another $1.5$~ms of free expansion, absorption images of the atoms along the $y$-axis are taken. Typical images are shown in Fig.~\ref{components}. These images allow for an analysis of the spatial structure and of the number of atoms in each spin component. Thus the longitudinal spin orientation can be determined.

\section{Spinor BEC in the Box-Potential}
\label{theory}

Figure~\ref{components} clearly shows an intriguing spatial structure of the clouds in the states $\ket{2,\pm 1}$. In the following, we develop a simple analytical model that allows for a detailed understanding of the observed spin excitation modes~\cite{Scherer2010} in the elliptical and the cylindrical trap.

\subsection{Spinor Hamiltonian}

The Hamiltonian of the system $\hat H$ is given by the sum of a single particle and an interaction term $\hat H =\hat H_0+\hat H_I$~\cite{Deuretzbacher2010}. The single particle Hamiltonian is 
\begin{equation}
\hat H_0= \int d^3r \sum_{m_F} \hat \psi_{m_F}^ \dag(\vec r) \left[-\frac{\hbar ^2 \Delta}{2M}+V (\vec r) -qm_F^2\right] \hat \psi_{m_F}(\vec r),
\end{equation}
where $\hat \psi_{m_F}(\vec r)$ is the field operator for each $m_F$ component, and the external potential is given by $V(\vec r)$. 
Since we are interested in spin-changing collisions only the quadratic Zeeman energy $-qm_F^2$ needs to be considered. The linear Zeeman effect does not contribute, since its net energy vanishes due to the conservation of the total spin orientation.

The interaction Hamiltonian is 
\begin{equation}
\hat H_I= \frac{1}{2}\int d^3r \sum_{\genfrac{}{}{0pt}{}{m_F,m_F^\prime,}{ m_f,m_f^\prime} } \hat \psi_{m_F}^\dag(\vec r) \psi_{m_F^\prime}^\dag(\vec r) U_{m_F,m_F^\prime}^{m_f,m_f^\prime} \hat \psi_{m_f}(\vec r)\hat \psi_{m_f^\prime}(\vec r)
\end{equation}
with the spin-dependent interaction strength $U_{m_F,m_F^\prime}^{m_f,m_f^\prime}\equiv U_0 \delta_{m_F,m_f}\delta_{m_F^\prime, m_f^\prime} +U_1 \vec f_{m_F m_f} \cdot \vec f_{m_F^\prime m_f^\prime}$ where $\vec f_{m_Fm_f^\prime}=(f^x_{m_Fm_f^\prime},f^y_{m_Fm_f^\prime},f^z_{m_Fm_f^\prime})^T$ and $f^{x,y,z}$ are the spin-1 Pauli matrices. The spin-dependent and the spin-independent coupling constants are given by $U_0=(7g_0+10g_2+18g_4)/35$ and $U_1=(-7g_0-5g_2+12g_4)/35$, where $g_F=4\pi\hbar a_F/M$ and $a_F$ is the $s$-wave scattering length for the channel with total spin $F$.

Since the BEC in our experiments is initially in the state $\ket{2,0}$, the dynamics of the spin states can be described with a spin Bogoliubov Ansatz
\begin{equation}
\hat\psi(\vec{r},t) = 
\left( \begin{pmatrix} 0 \\0 \\ \sqrt{n_0(\vec{r})} \\ 0 \\ 0 \end{pmatrix} + 
\begin{pmatrix}  \delta\hat\psi_{-2} \\  \delta\hat\psi_{-1} \\  \delta\hat\psi_{0} \\  \delta\hat\psi_{1} \\  \delta\hat\psi_2 \end{pmatrix}  
\right) e^{-i \mu t},
\end{equation}
where the BEC in the state $\ket{2,0}$ is described as a classical field with the chemical potential $\mu$ and the field operators $\delta \hat{\psi}_{m_F}$ for small fluctuations of each spin state. The population in the states $\ket{2,\pm 2}$ can be neglected for the short spin dynamics times, since the probability of spin changing collisions to these states is small due to small Clebsch-Gordan coefficients. 

The resulting Hamiltonian, up to second order in $\delta \hat{\psi}_{\pm 1}$, is given by
\begin{align}
\label{H}
\hat{H}=&\int d^3r\sum_{m_F=\pm1}\delta \hat{\psi}^{\dag}_{m_F} [\hat{H}_\text{eff} + q] \delta \hat{\psi}_{m_F}\\ &+\Omega (\vec{r}) \left [\delta \hat{\psi}^{\dag}_1 \delta \hat{\psi}^{\dag}_{-1} + \delta \hat{\psi}_1 \delta \hat{\psi}_{-1}\right ] \nonumber
\end{align}
where $q$ represents the quadratic Zeeman energy. The term preceded by $\Omega (\vec{r})=U_1n_0(\vec{r})$ accounts for the spin changing collisions and $\hat{H}_\text{eff}$ represents the effective single particle Hamiltonian 
\begin{equation}
\hat{H}_\text{eff}=-\frac{-\hbar^2\nabla^2}{2m} + V(\vec{r})+(U_0+U_1)n_0(\vec{r})-\mu.
\end{equation}
for the atoms transferred to the state $\ket{2,\pm 1}$. Hence these atoms experience an effective trapping potential given by
\begin{equation}
V_\text{eff}(\vec{r})=V(\vec{r})+(U_0+U_1)n_0(\vec{r})-\mu.
\end{equation}
\label{Veff}

Based on this Hamiltonian a full numerical analysis of the experiments is possible~\cite{Scherer2010}. However, to gain an insight in the underlying physical processes it is advantageous to make a number of simplifying assumptions discussed in the following.

\subsection{Effective box potential}

A closer look at the effective potential $V_\text{eff}$ allows for several simplifying assumptions. These approximations result in an analytically solvable single-particle Hamiltonian and thus allow for a deep insight in the underlying processes. 

In the Thomas-Fermi approximation the density distribution of the BEC mimics the shape of the trapping potential $n_0(\vec{r})\approx U_0^{-1}(\mu-V(\vec{r}))$. Hence within the Thomas-Fermi radius $r<\rtf$ the atoms in the states $\ket{2,\pm 1}$ experience a flat bottomed potential, which is modified by the term $U_1 n_0(\vec{r})$, corresponding to a small parabolic repulsion.  In the presented analysis, this term is neglected, since it typically has a height of $\approx h\times 30$~Hz. 

\begin{figure}[ht]
\centering
\includegraphics{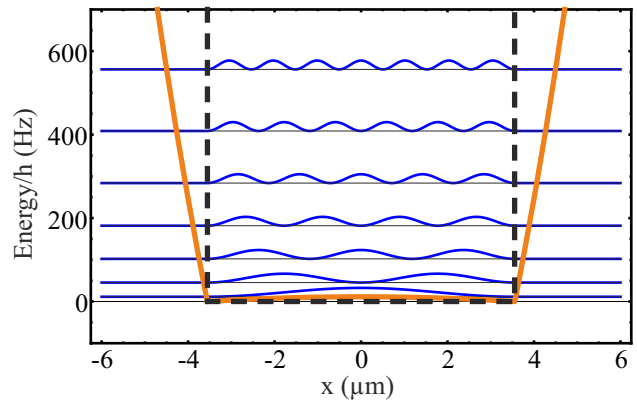}
\caption{Sketch of the effective potential for atoms in the states $\ket{2,\pm 1}$ and its approximation under typical experimental conditions. The harmonic external potential is modified by the repulsive interaction with the atoms in the state $\ket{2,0}$ . The resulting effective potential (orange) is approximated by a simple box potential (dashed, black). In addition, the density distributions of the eigenstates of the 1D box potential are shown at the  position of their eigenenergies (blue).}
\label{boxpotential}
\end{figure}

Outside the Thomas-Fermi radius, the potential is given by the harmonic confinement of the dipole trap $V(\vec{r})=m/2\sum\omega_i^2 x_i^2$, which rises sharply at $\rtf$. In a further step of simplification, this confinement is approximated by infinite walls, $V_\text{eff}=\infty$ for $r>\rtf$. Within these approximations, the process can hence be analyzed in a simple box potential with the size of the Thomas-Fermi radius. This simplified situation is shown in Fig.~\ref{boxpotential} for the one-dimensional case.

The choice of trapping configuration allows for another simplification of the three-dimensional problem. The cylindrical trap configuration has two strong radial trap frequencies of nearly the same size ($\omega_x=$187 Hz, $\omega_z=$183 Hz) and a considerably weaker axial trap frequency ($\omega_y=$65 Hz). This indicates that the analysis can be limited to a two dimensional cylindrical box potential to evaluate the radial spin excitation modes. 

\subsection{Spinor dynamics in the one-dimensional box potential}

To provide an insight in the spin dynamics process we first restrict ourselves to the simplified one-dimensional box potential introduced above. This allows for a qualitative analysis of the spatial structure along the principle axis of an elliptical trap. 

The spin excitation modes in the  one-dimensional box potential can be analyzed by evaluating the single particle eigenfunctions of the effective Hamiltonian.

In this case the well known solutions are plane waves with a discrete wave vector $k_n=n \pi/(2\ \rtf)$ of the form
\begin{equation}
\varphi_{n}(x)= \frac{1}{\sqrt{\rtf}} \sin(k_n x+n\frac{\pi}{2})
\end{equation}
with eigenenergies $\epsilon_n=\hbar^2 k_n^2/(2m)$. The density distribution hence consists of a chain of neighboring maxima, and the number of these maxima is given by $n$ as shown in Fig.~\ref{boxpotential}. 

To analyze the stability of the excitation modes in this system, we expand the Hamiltonian using $\delta \hat \psi_{m_F}=\sum_n \hat{a}_{n,m_F} \varphi_n(x)$ to obtain $\hat H= \sum_n \hat H_n$ with
\begin{align}
\label{H_n}
\hat H_n =& (\epsilon_n + q) \sum_{m_F} \hat{a}\dag_{n,m_F} \hat{a}_{n,m_F}\\
&+ \Omega \left(\hat a\dag_{n,1} \hat a\dag_{n,-1} + \hat{a}_{n,1} \hat a_{n,-1} \right). \nonumber
\end{align}

Thus the Heisenberg equation for each mode $i \hbar\frac{d}{dt} \hat{a}_{n,m_F}^{(\dag)} =[ \hat{a}_{n,m_F}^{(\dag)}, \hat H]$ can be represented by
\begin{equation}
\label{TE_a}
i \hbar\frac{d}{dt} 
\begin{pmatrix}
\hat a_{n,1}\\
\hat{a}_{n,-1}^\dag
\end{pmatrix} = 
\begin{pmatrix}
\epsilon_n+q & \Omega \\
-\Omega & -\epsilon_n-q 
\end{pmatrix}
\begin{pmatrix}
\hat a_{n,1}\\
\hat a_{n,-1}^\dag
\end{pmatrix}
\end{equation}

and the time evolution of the system can be obtained from the eigenvalues $\xi_{n}=\sqrt{(\epsilon_{n} + q )^2 - \Omega^2}$ of this matrix. Excitation modes with real eigenvalues are stable, whereas imaginary eigenvalues lead to an exponential amplification of the population of the mode $\varphi_{n}$. A particularly interesting behavior arises when the eigenenergy of the effective Hamiltonian is equal to the quadratic Zeeman energy $\epsilon_{n}+q=0$. In this case the imaginary eigenvalue $\xi_{n}= i |\Omega|$ of (\ref{H_n}) reaches a maximum and the corresponding mode is maximally unstable, causing resonances in spin dynamics~\cite{Klempt2009}.

\begin{figure}[ht]
\centering
\includegraphics{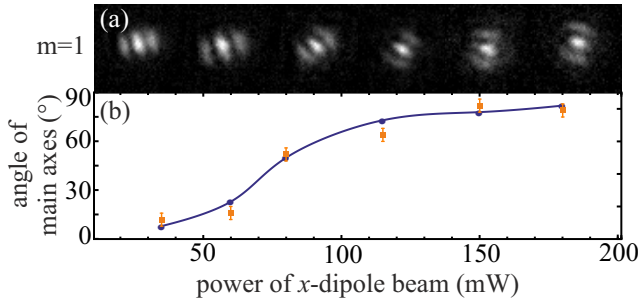}
\caption{Orientation of the spin excitation mode in a rotated elliptical trap. (a) Time-of-flight images of the spatial distribution of atoms in the state $\ket{2,1}$ as a function of the power $P_x$ in a configuration with a deliberate misalignment between the two trapping beams. (b) The angles of the distributions (orange squares) are shown in comparison with the angle of the strongest trap axis obtained from a trap frequency measurement (blue dots, solid line).}
\label{fighters}
\end{figure}

Figure~\ref{fighters} displays the density distribution of a spin excitation mode for different orientations of an elliptical trap. The agreement between the orientation of the trap and the spin excitation mode clearly shows that the observed spatial distributions are indeed a feature of the trapping potential. The agreement between the predicted and the observed shape of the distribution along the principal axes also verifies that the one-dimensional description of the system is valid. Note however, that this one-dimensional  description naturally fails to account for the two-dimensional shape of the excitation modes and their dependence on the trap configuration.

\subsection{Spinor dynamics in the cylindrical trap}
\label{spinorcylinder}

\begin{figure}[ht]
\centering
\includegraphics{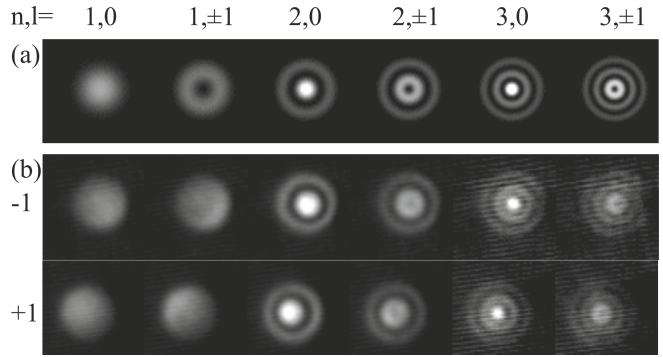}
\caption{Density distributions of spin excitation modes in a cylindrical trap. (a) Expected profiles in a cylindrical box potential according to Eq.~(\ref{Wave2D}). (b) Images of experimental density profiles recorded in time-of-flight, averaged over 30 realizations.}
\label{modes}
\end{figure}

The analysis presented above can be extended to the physically richer case of the cylindrical trap configuration. For a two-dimensional cylindrical box, the single-particle eigenfunctions are given by
\begin{equation}
\label{Wave2D}
\varphi_{n,l}(r,\gamma)=\frac{1}{\sqrt{\pi} \rtf J_{l+1}\left (\beta_{n,l}\right )}
J_l\left (\beta_{n,l}\frac{r}{\rtf}\right )e^{il\gamma}
\end{equation}
with corresponding eigenenergies $\epsilon_{n,l}=\hbar^2\beta^2_{n,l}/2m\rtf^2$. Here, $J_l$ are Bessel functions of the first kind and $\beta_{n,l}$ is the $n$'th zero of $J_l$. The modes can be identified by the quantum numbers $n$ for the radial excitations and $l$ for the angular momentum along the y-direction. Figure~\ref{modes}~(a) shows plots of the corresponding single-particle density distributions.

Similar to the previous case, the Hamiltonian can be expanded in these eigenfunctions $\delta \hat{\psi}_{m_F} (\vec{r}) = \sum_{n,l} \hat{a}_{n,l,m_F} \varphi_{n,l} (\vec{r})$. One obtains $\hat{H} =\sum_{n,l} \hat{H}_{n,l}$, where $\hat{H}_{n,l}$ is given by
\begin{align}
\label{H_nl}
\hat{H}_{n,l}=&\left(\epsilon_{n,l} + q\right) \sum_{m_F} \hat{a}^\dag_{n,l,m_F} \hat{a}_{n,l,m_F}\\ \nonumber
&+ \Omega \left( \hat{a}^\dag_{n,l,1} \hat{a}^\dag_{n,-l,-1} + \hat{a}_{n,l,1} \hat{a}_{n,-l,-1} \right).
\end{align}
Note that the counter-rotating modes $H_{n,l}$ and $H_{n,-l}$ are energetically degenerate since $\epsilon_{n,l}=\epsilon_{n,-l}$.

The Heisenberg equation for the creation and annihilation operators is analogous to Eq.~(\ref{TE_a}) and excitation modes with real eigenvalues $\xi_{n,l}=\sqrt{(\epsilon_{n,l} + q )^2 - \Omega^2}$ are stable, whereas imaginary eigenvalues lead to an exponential amplification of the population in the mode $\varphi_{n,l}$ with the instability rate $\text{Im}(\xi_{n,l})/h$. 

\subsubsection{Magnetic field position of spin excitation modes}
\label{modeposition}

\begin{figure}[t]
\centering
\includegraphics{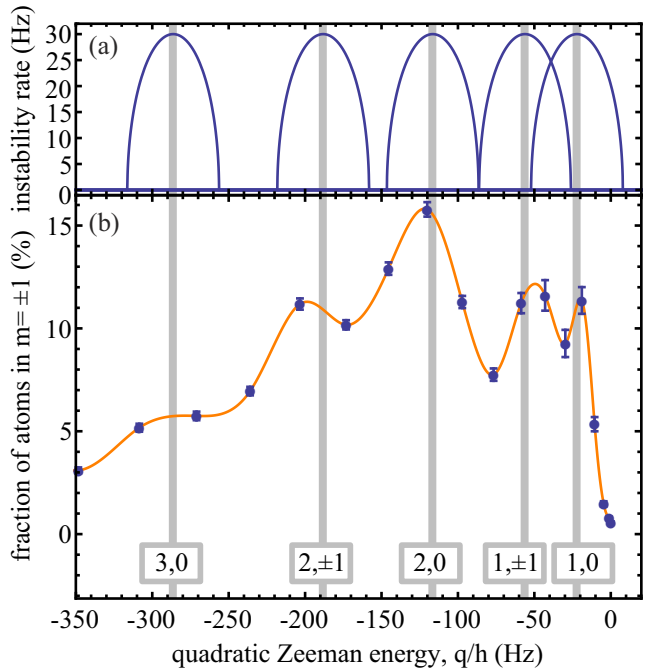}
\caption{Spin dynamics rate in a cylindrical trap as a function of the quadratic Zeeman energy. (a) Instability rate calculated for the cylindrical box potential. (b) Measured number of atoms in the states $\ket{2,\pm 1}$ after a spin dynamics time of 17~ms. The orange line is a guide to the eye. The vertical gray bars indicate the maxima of the instability rate with the corresponding quantum numbers $(n,l)$.}
\label{stability}
\end{figure}

Figure~\ref{stability} compares the observed spin dynamics resonances in the cylindrical trap with the theoretically expected energies of maximal instability in a cylindrical box potential. The experiment is conducted as outlined in section~\ref{spinexperiment} with a spin dynamics time of 17~ms. Several resonances in the transfer efficiency are clearly visible, indicating maxima of the instability rate. To identify the quantum numbers of the corresponding excitation modes, only the Thomas-Fermi radius was varied in the calculation of the eigenenergies $\epsilon_{n,l}$ to fit the maxima in Fig.~\ref{stability}~(a) to the observed maxima of the instability rate in Fig.~\ref{stability}~(b). The resulting value of $\rtf=3.9\mu$m, is in very good agreement with the value of $3.7\mu$m obtained from a mean field calculation. The small discrepancy probably reflects the assumption of infinitely high box walls. The spatial distribution observed on these resonances is shown in Fig.~\ref{modes}~(b), clearly showing that the expected shape of the excitation modes can be observed in the expanded density profiles after TOF absorption imaging.

The images shown in Fig.~\ref{modes} also allow for a visual identification of the quantum numbers. The number of maxima of the density along the radius corresponds to the quantum number $n$. The second quantum number $l$ indicates the angular momentum of the modes. Modes with $l=\pm 1$ form a vortex which results in a density minimum at the center of the clouds.

This analysis clarifies the origin of the spin excitation resonances in the cylindrical trap. The good agreement shown in Fig.~\ref{modes} and Fig.~\ref{stability} verifies that a two dimensional analysis in a simple cylindrical box potential is justified and allows for the identification of the observed excitation modes. The model does however not give correct instability rates, primarily because it does not include the mode overlap between the BEC and the spin excitation modes~\cite{Klempt2010}.
 
\subsubsection{Analysis of spin excitation mode contribution}
\label{modecontribution}

\begin{figure}[ht]
\centering
\includegraphics*[width=0.85\columnwidth]{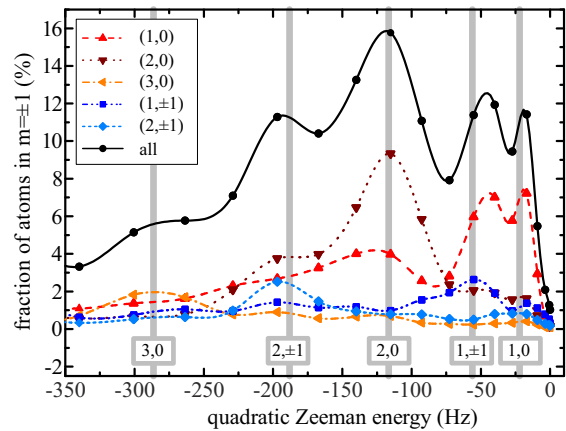}
\caption{Contribution of the excitation modes to the fraction of atoms transferred to the state $\ket{2,\pm 1}$. The black dots indicate the experimental data shown in Fig.~\ref{stability}. The weights of the contributing excitation modes according to our fit are shown as triangles and diamonds (see legend). The vertical gray bars show the result of the analysis in section~\ref{modeposition}. All lines are guides to the eye.}
\label{Besselfits}
\end{figure}

The method to identify the quantum numbers presented above is applicable when the modes are well separated. This is the case close to the maxima of the observed spin dynamics rate, but it fails when a superposition of several modes is excited. Therefore a second method to analyze the density profiles was used, which allows us to identify the contributing excitation modes at each energy. This is achieved by fitting a superposition of the density distributions of the excitation modes ($n=1$ to $n=3$) to the observed averaged density profiles. Each fit reveals the contributions of the individual excitation modes. The results are shown in Fig.~\ref{Besselfits}. 

Each resonance clearly shows a strong contribution of the appropriate spatial mode. On the first resonance at $\approx -22$~Hz, the $(n=1,l=0)\equiv(1,0)$ contribution is the strongest. The second resonance at $\approx -50$~Hz is a superposition of the $(1,0)$ and $(1,\pm1)$ modes while the third resonance at $\approx -120$~Hz is dominated by the $(2,0)$ mode. Similarly the fourth and fifth resonance at $\approx -180$~Hz and $\approx -280$~Hz have strong contributions of the associated $(2,\pm1)$ and $(3,0)$ modes. The difference between the ideal Bessel modes and the experimentally measured distributions lead to spurious contributions of other (lower lying) spatial modes. In addition, imperfections of the imaging system and the low signal-to-noise ratio add to the weight of these spurious contributions. To overcome these problems, pattern recognition algorithms~\cite{Cordes2011} or image processing techniques~\cite{Ockeloen2010} might be employed. Nonetheless our simple model correctly identifies the dominant spatial structure at the resonance positions, showing that the effective box potential approach is well justified.

\section{Symmetry breaking in spinor BEC}
\label{SymmetryBreaking}

The spin excitation modes in a cylindrical potential allow for the observation of a spatial- and a spin-symmetry breaking process~\cite{Scherer2010}. Both of these processes can be analyzed within the framework of the cylindrical box potential presented above.

Let us consider the case of a spin excitation mode with non-vanishing angular momentum $(l\neq 0)$. In this case, two degenerate modes are present in the system, one mode with positive angular momentum $l=+|l|$ rotating clockwise (vortex) and one with negative angular momentum $l=-|l|$ rotating counterclockwise (antivortex). If just one mode is populated by spinor dynamics, the resulting density distribution will be cylindrically symmetric. However, if superpositions of vortex and antivortex modes are populated, they interfere and form an azimuthal standing wave which is no longer cylindrically symmetric and shows a clear orientation. 

\begin{figure}[h]
\centering
\includegraphics{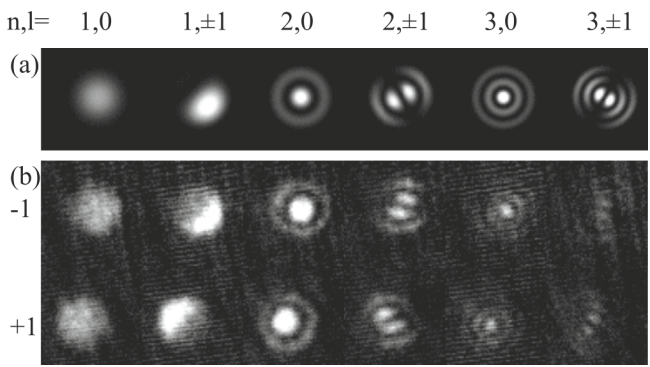}
\caption{Density distributions of individual spin excitations in a cylindrical trap. (a) Calculated density distributions for a superposition of the vortex-antivortex modes $(n,\pm 1)$ (where applicable). In the case of the mode $(1,\pm 1)$, an admixture of the neighboring mode $(1,0)$ was included. (b) Experimental absorption images of individual spin excitations.}
\label{modesymmetry}
\end{figure}

Figure~\ref{modesymmetry}~(a) shows calculated density distributions of such superposition states based on Eq.~(\ref{Wave2D}) for excitation modes with quantum numbers $(2,\pm 1)$ and $(3,\pm 1)$. These distributions assume equal populations of the vortex and antivortex modes and a fixed overall phase $\phi_{n,l,m_F}$~\footnote{Note that an additional index $m_F$ indicates that the phase can differ for $m_F$ states.} was randomly assigned to each mode. Since the pattern is given by
\begin{eqnarray}
\label{eq:phase}
\left| \langle\delta \hat\psi_{m_F}(r,\gamma)\rangle\right|^2 = \left|\sum_{l=\pm |l|}\langle\hat a_{n,l,m_F}\rangle\varphi_{n,l}(r,\gamma)\right|^2 \nonumber\\
\propto  1+(-1)^{|l|}\cos(\phi_{n,l,m_F}- \phi_{n,-l,m_F}+2 |l| \gamma),
\end{eqnarray}
the spatial orientation is determined by the difference between the overall phases, and the term $2 |l| \gamma$ corresponds to the azimuthal standing wave. 

In the experimental case, the excitation modes for $n>1$ are populated due to the parametric amplification of vacuum fluctuations~\cite{Klempt2010}. Therefore the phases $\phi_{n,l,m_F}$ are chosen arbitrarily by the system, and hence the angle of the density distribution is expected to be different for each experimental realization, corresponding to spontaneous spatial symmetry breaking. Figure~\ref{modesymmetry}~(b) shows images of individual experimental realizations to illustrate this behavior.

Besides the breaking of the spatial symmetry, which occurs in each $m_F$ state individually, the local spin symmetry can also be broken. Since all $m_F$ components are confined by the same trap, a difference in the orientation of the two clouds in the states $\ket{2,\pm 1}$ corresponds to a spatially varying local longitudinal spin orientation. This case can be observed on the mode $(2,\pm 1)$ as shown in Fig.~\ref{modesymmetry}~(b). It shows that the phases $\phi_{n,l,m_F}$ must differ in the states $\ket{2,\pm 1}$, and hence an analysis of these phase correlations is required.

\subsection{Theoretical symmetry breaking analysis}

To investigate the symmetry breaking process, an analysis of the phase correlations in the states $\ket{2,\pm 1}$ is required. We first calculate the states which are generated on the unstable modes ($\xi_{n,l}=i|\xi_{n,l}|$) in the two-mode Fock-basis $\ket{n_{m_F=-1}}\ket{n_{m_F=+1}}\equiv \ket{n_{-1},n_{+1}}_F$, where $n_{\pm 1}$ is the number of atoms in the Zeeman-states. These states allow for a calculation of the phase states as defined in Ref.~\cite{Barnett1990}, which provide the phase correlations of interest. 

The time evolution operator of the spin dynamics process is given by
\begin{equation*}
\hat U(t) \equiv\exp(-i \hat H_{n,l}t/\hbar).
\end{equation*}
On a spin excitation resonance ($\epsilon_{n,l}+q=0$), this operator corresponds to the well known two-mode squeezing operator with a squeezing parameter $\zeta\equiv\frac{\Omega t}{\hbar}e^{-i \pi/2}$ and can thus be written as 
\begin{equation*}
\hat U_R(t)=\exp\left[\zeta \left( \hat{a}^\dag_{n,l,1} \hat{a}^\dag_{n,-l,-1} + \hat{a}_{n,l,1} \hat{a}_{n,-l,-1} \right) \right].
\end{equation*}

To obtain an explicit form of the time evolution operator $\hat U(t)$, we solve the Heisenberg equation by diagonalizing the Hamiltonian $\hat H_{n,l}$. This is achieved by introducing the quadratures 
\begin{equation*}
\underbrace{
\begin{pmatrix}
\hat X^{(1)}_{n,l,m_F}\\
\hat X^{(2)}_{n,-l,-m_F}
\end{pmatrix}}_{\equiv \vec{\hat X}}=
\underbrace{
\frac{1}{\sqrt{2 \sin(2\theta_{n,l})}}
\begin{pmatrix}
e^{i\theta_{n,l}} & e^{-i\theta_{n,l}}\\
e^{-i\theta_{n,l}}& e^{i\theta_{n,l}}
\end{pmatrix}
}_{\equiv A}
\underbrace{
\begin{pmatrix}
\hat a_{n,l,1}\\
\hat a_{n,-l,-1}^\dag
\end{pmatrix}}_{\equiv \vec{\hat a}}.
\end{equation*}
where $\cos 2 \theta_{n,l}=(\epsilon_{n,|l|}+q)/\Omega$. 

Thus the Hamiltonian (\ref{H_nl}) can be rewritten in the form
\begin{equation*}
\hat H_{n,l}\! =\! \frac{|\xi_{n,l}|}{2} \sum_{m_F,l} \left[ \hat X^{(1)}_{n,l,m_F} \hat X^{(2)}_{n,l,m_F}+\hat X^{(2)}_{n,l,m_F} \hat X^{(1)}_{n,l,m_F} \right]
\end{equation*}

The Heisenberg equations for the quadratures are $i\hbar \frac{d}{dt}\hat X^{(1,2)}_{n,l,m_F}=[X^{(1,2)}_{n,l,m_F}\hat H_{n,l}]=i\xi_{n,l}X^{(1,2)}_{n,l,m_F}$ and their time evolution is
\begin{equation}
\begin{pmatrix}
\hat X^{(1)}_{n,l,m_F}(t)\\
\hat X^{(2)}_{n,-l,-m_F}(t)
\end{pmatrix}=
\underbrace{
\begin{pmatrix}
e^{\xi_{n,l}t} & 0\\
0& e^{-\xi_{n,l}t}
\end{pmatrix}}_{\equiv T}
\begin{pmatrix}
\hat X^{(1)}_{n,l,m_F}(0)\\
\hat X^{(2)}_{n,-l,-m_F}(0)
\end{pmatrix}.
\end{equation}

Based on these solutions the time evolution of the creation and annihilation operators are found by using the relation $\vec{\hat a}(t)=A^{-1}TA~\vec{\hat a}(0)\equiv \hat U_{n,l}(t)\vec{\hat a}(0)$. 

\begin{align}
\hat{U}_{n,l}(t)=
\begin{pmatrix}
U_{n,l} & \tilde U_{n,l}\\
\tilde U_{n,l}^* &  U_{n,l}^*
\end{pmatrix}
\end{align}
where
\begin{align}
U_{n,l}&=\frac{-i}{\sin (2\theta_{n,l})}\sinh\left(\frac{\xi_{n,l}t}{\hbar}+2\theta_{n,l}\right)\\
\tilde U_{n,l}&= \frac{-i}{\sin (2\theta_{n,l})} \sinh\left(\frac{\xi_{n,l}t}{\hbar}\right).
\end{align}
To obtain the state $\ket{\zeta}=\hat U_{n,l}(t)\ket{0,0}_F$ generated by parametric amplification of the two-mode vacuum state, we use the fact that the vacuum state is an eigenstate of the annihilation operator with eigenvalue zero $\hat a_{n,l,m_F}(t)\ket{0,0}_F=0$. By multiplying this equation with the time evolution operator and by using its unitarity, we obtain the eigenvalue equation 
\begin{align*}
&\hat U_{n,l}(t) \hat a_{n,l,m_F} \hat U_{n,l}^\dag(t) \hat U_{n,l}(t) \ket{0,0}_F \nonumber\\
&=\left( U_{n,l}(-t) \hat a_{n,l,m_F}(0) + \tilde U_{n,l}(-t) \hat a_{n,-l,-m_F}^\dag(0) \right) \ket{\zeta}=0,
\end{align*}
expanding the state in terms of two-mode Fock-states $\ket{\zeta}=\sum_{k,k}c_k\ket{k,k}_F$ \cite{GerryKnight}, we find
\begin{align*}
\ket{\zeta}=c_{0}\sum_k \left(-\frac{\tilde U_{n,l}(-t)}{U_{n,l}(-t)} \right)^k \ket{k,k}_F.
\end{align*}
In this equation, the sum adds states with exactly the same number of particles in the two modes, so-called twin-Fock states. This results in an equal number of atoms in the two different spin states in every realization. The coefficient $c_{0}$ is obtained from the normalization condition. If we assume that the system is on one of the resonances ($\epsilon_{n,l}+q=0$), with negligible contribution from other resonances, the state has the form
\begin{equation}
\label{2modeSqueezedVacuum}
\ket{\zeta}=\frac{1}{\cosh(\xi_{n,l}t/\hbar)}\sum_k (-i)^k \tanh^k \left(\frac{\xi_{n,l}t}{\hbar}\right)\ket{k,k}_F.
\end{equation}
This exactly corresponds to the two-mode squeezed vacuum-state with squeezed and antisqueezed two-mode quadratures and equal particle numbers.

In the context of symmetry breaking, our main interest are the phase correlations between the modes in the two states. These correlations were analyzed in Ref.~\cite{Barnett1990} for the state given in Eq.~(\ref{2modeSqueezedVacuum}). It was shown that the expectation value of the phase sum is constant $\langle \phi_p + \phi_q \rangle = -\pi/2 $ and that its variance is given by $\Delta (\phi_p + \phi_q)=\frac{\pi^2}{3}+4\ \text{dilog}\left[ 1+\tanh\left( \frac{\xi_{n,l}t}{\hbar}\right) \right] $, where $\text{dilog}[~]$ is the dilogarithm function. The variance hence tends to zero, $\lim_{t\rightarrow \infty}\Delta (\phi_p + \phi_q)=0$. Thus the phase sum of the two modes is squeezed with increasing time.

These results allow for an interpretation of the spatial symmetry breaking in terms of the squeezing of the phase sum. Let us initially assume that only two degenerate vortex- and antivortex modes e.g. the modes $(2,\pm 1)$ are populated by spin dynamics. In this case, one obtains a twofold two-mode squeezing, where the two phase sums are equal. The phase sums of the vortex mode of the state $\ket{2,1}$ and of the antivortex mode in the state $\ket{2,-1}$ are squeezed 
$\langle \phi_{n,l,1}+\phi_{n,-l,-1} \rangle= -\pi/2$ and vice versa $\langle \phi_{n,-l,1}+ \phi_{n,l,-1} \rangle= -\pi/2$. Hence it follows for large squeezing-factors $|\zeta|$, that the phase-differences in the states $\ket{2,\pm 1}$ are equal, $\phi_{n,l,m_F}- \phi_{n,-l,m_F}=\phi_{n,l,-m_F}- \phi_{n,-l,-m_F}$. Therefore the spatial orientation of the density distributions in both states is equal (see Eq.~(\ref{eq:phase})) and a breaking of the spatial symmetry is expected, but not a breaking of the longitudinal spin symmetry.

Besides the resonance, the squeezing factor $|\zeta|$ and thus the phase sum correlation gets smaller and the phase-differences in the states $\ket{2,\pm 1}$ are not necessarily the same. Hence the probability of observing symmetry breaking in the local longitudinal spin increases due to the different spatial orientation of the density distributions of the two clouds. Thus local spin symmetry is only observed if the squeezing factor is high enough.

\subsection{Experimental analysis of symmetry breaking on the  mode $(2,\pm 1)$} 

The most striking experimental results were obtained on the resonance $(2,\pm 1)$. While the distribution is symmetric on the averaged images in Fig.~\ref{modes}, the distribution observed in individual experimental realizations (Figure~\ref{modesymmetry}~(b)) clearly shows both spatial and spin symmetry breaking. Similarly, it is possible to observe breaking of both symmetries on the resonance $(3,\pm 1)$, but the signal-to-noise ratio is typically insufficient for a quantitative analysis.

We also observe both types of symmetry breaking on the resonance $(1,\pm 1)$, as shown in Fig.~\ref{modesymmetry}. However, in this case the density distributions are due to superpositions of several modes with and without angular momentum, since the instability rates of neighboring modes $(1,0)$ and $(2,0)$ are large. This is confirmed by the fact that the shape of the density distribution on this resonance is not symmetric in the averaged images in Fig.~\ref{modes}. We therefore conclude that the symmetry breaking on this resonance is classical, caused by experimental imperfections such as magnetic field gradients.

Within the following quantitative analysis we therefore focus on the symmetry breaking of the resonance $(2,\pm 1)$. Within this analysis the orientations of the individual clouds and their distribution have to be determined. For this purpose two independent methods are used and only the images where both methods agree within an error interval are used. This procedure allows for the exclusion of images, where the intrinsic number of transfered atoms is too small to obtain its orientation.

\begin{figure}
\centering
\includegraphics{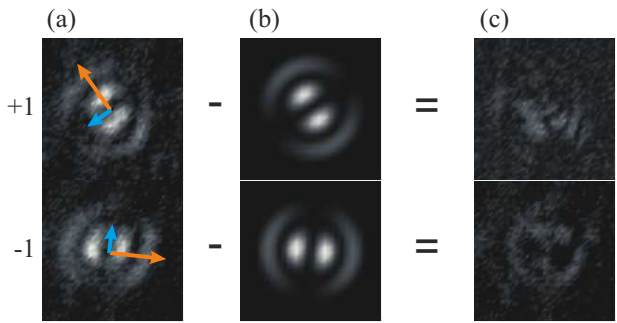}
\caption{Measurement of the orientation of the density distributions. (a) The measured density distributions are shown together with the resulting eigenvectors of the quadrupole tensor. The orientation of the biggest eigenvector (orange arrow) corresponds to the angle of the density distribution. (b) Result of fits to the measured distributions using an equal superposition of the modes $(2,\pm 1)$. (c) Absolute value of the difference between the measured and the fitted density distributions.}
\label{fitmethods}
\end{figure}

\begin{figure}
\centering
\includegraphics*[width=0.85\columnwidth]{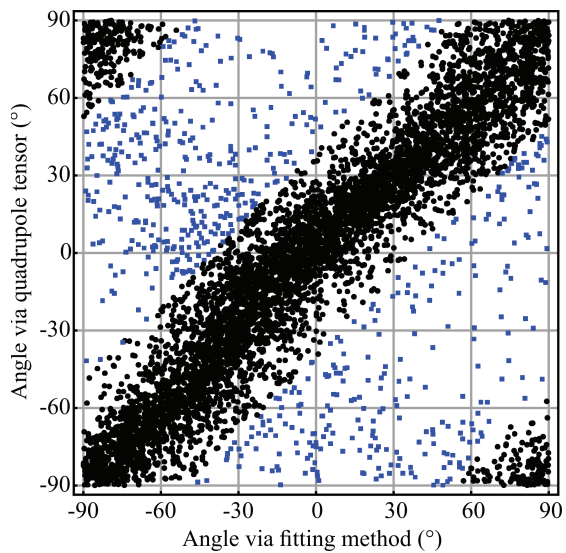}
\caption{Comparison of the two methods to determine the orientation of the density distributions. Only experimental results where both methods agree within $40^\circ$ (black dots) where used for further analysis.}
\label{Analyzeangles}
\end{figure}

The first method fits the density distribution expected for a superposition of the vortex and antivortex modes to each experimental density distribution (see Fig.~\ref{fitmethods} (b)). The fit directly yields the distribution's orientation relative to the fixed camera axis. In the second method, we calculate the two-dimensional quadrupole tensor for each image
\begin{equation}
Q_{i,j}=\sum_k  n_{k} \left(3(\vec r_{k})_i (\vec r_{k})_j - \delta_{i,j} \vec r_{k} \right),
\end{equation}
where $i,j \in \left\{x,y \right\}$ represent the two spatial dimensions and we sum over all pixels of the image. The atomic densities on the pixels are $n_k$ and the position vectors are $r_k$, where the origin is chosen at the center of mass in each individual cloud. The eigenvectors of this tensor give the principal axes of the density profile (see Fig.~\ref{fitmethods} (a)). For sufficient data quality, the calculated orientation should be the same as obtained from the fitting method.

Figure~\ref{Analyzeangles} shows the correlation between the angles obtained by the two methods. The difference of the angles has a standard deviation of $15^\circ$. For the analysis of the distribution, only the images where both methods agree within $40^\circ$ were taken into account, corresponding to $78\%$ of the measurements. We have verified that the experimental results were stable under variation of this interval.

\begin{figure}
\centering
\includegraphics{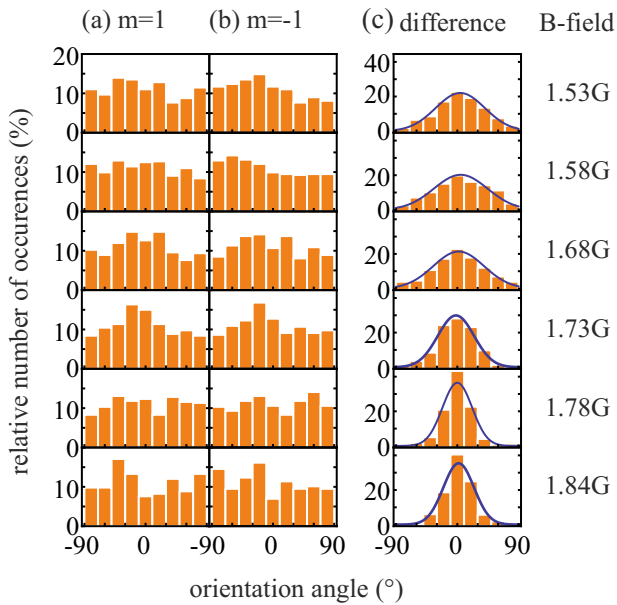}
\caption{Orientation of individual density distributions recorded in the vicinity of the mode $(2,\pm 1)$. (a), (b) Distribution of angles for the states $\ket{2,1}$ and $\ket{2,-1}$. (c) Distribution of the difference between the angles of orientation.}
\label{angles}
\end{figure}

To measure the varying degree of spatial and spin symmetry breaking, the orientations of the density distributions in the states $\ket{2,\pm 1}$ were recorded at different magnetic fields around the resonance $(2,\pm 1)$. Figure~\ref{angles} shows the distribution of these orientations for each state and their difference for six different magnetic fields. 

Since no significant preferred orientation is observed in the spatial distributions of the individual clouds, the cylindrical symmetry is indeed broken spontaneously. This also confirms that the symmetry breaking is not induced by the remaining asymmetry of the trapping potential, by a spurious production of seed atoms or by magnetic-field gradients. 

Moreover, the distributions of the relative angle between the two clouds reveal the second symmetry breaking effect. The distribution is peaked around $0^\circ$ for all magnetic fields showing that the two angles are correlated as expected by theory. At a field of $1.78$G the width of the distribution is smallest, matching the resolution of the applied angle measurement. In this case, the local spin of the system remains $0$, indicating that the squeezing parameter and thus the phase-sum squeezing of the degenerate vortex- and antivortex modes is maximal.

\begin{figure}[ht]
\centering
\includegraphics*[width=0.85\columnwidth]{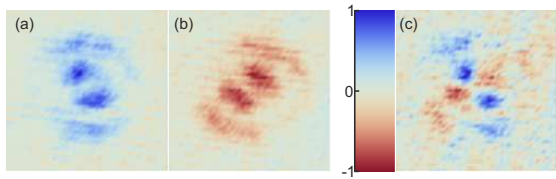}
\caption{Visualisation of the doubly broken symmetry. (a),(b) Spin density distribution of the individual components after time-of-flight imaging. (c) Resulting spin pattern obtained by subtracting the distributions.}
\label{fig:SpinPattern}
\end{figure} 

At higher or lower magnetic fields however the squeezing parameter decreases and thus differing orientations in the states $\ket{2,\pm 1}$ become more probable and the distributions broaden. In this case, both the spatial symmetry and the local longitudinal spin symmetry are broken. Figure~\ref{fig:SpinPattern} explicitly shows the doubly broken symmetry by subtracting the normalized spin density  distributions in the states $\ket{2,\pm 1}$ after TOF. This leads to an intricate spin pattern shown in Fig.~\ref{fig:SpinPattern}~(c) that reflects the spin pattern in the trap before the TOF imaging sequence. The local spin clearly varies over the cloud and thus the initial homogeneous spin distribution is spontaneously broken. 

\section{Conclusion}
\label{conclusion}

In summary we have presented a strikingly simple analytical model for the theoretical analysis of spin dynamics and symmetry breaking in a spinor condensate. This allows us to obtain an excellent intuitive understanding of the process and provides good quantitative agreement with experimental results.
 
A detailed description of the experimental techniques used to prepare, investigate and detect spinor BECs is given. This justifies a model which approximates the effective trapping potential for the atoms produced in the states $\ket{2,\pm 1}$ with a cylindrical box potential. Within this potential the observed shape of the spin excitations and their resonance positions can easily be understood. This provides the basis for an understanding of the spontaneous symmetry breaking of the density distributions and of the longitudinal spin orientation. In particular the superposition of vortex-antivortex modes with opposite angular momentum and quantum fluctuations of the relative phases lead to the symmetry breaking processes.

Our results show that spinor gases constitute an exceptionally suitable system for the detailed analysis of symmetry breaking, and  its close connection to multimode squeezing during parametric amplification. This allows for applications of the process to produce correlated quantum states for atom interferometry below the shot noise limit~\cite{Luecke2011,Gross2011}.

\section{Acknowledgments}

 We acknowledge support from the Centre for Quantum Engineering and Space-Time Research QUEST and from the Deutsche Forschungsgemeinschaft (Research Training Group 1729). We also thank the Danish Council for Independent Research, and the Lundbeck Foundation for support.

\bibliography{Scherer}

\end{document}